# Superconductivity in Few-Layer Stanene

Menghan Liao[1#], Yunyi Zang[1#], Zhaoyong Guan[1], Haiwei Li[1], Yan Gong[1], Kejing Zhu[1], Xiao-Peng Hu[1,2], Ding Zhang[1,2 *], Yong Xu[1,2,3*], Ya-Yu Wang[1,2], Ke He[1,2], Xu-Cun Ma[1,2], Shou-Cheng Zhang[4], and Qi-Kun Xue[1,2*]

[1]State Key Laboratory of Low-Dimensional Quantum Physics, Department of Physics, Tsinghua University, Beijing, 100084, China

[2]Collaborative Innovation Center of Quantum Matter, Beijing, China

[3]RIKEN Center for Emergent Matter Science (CEMS), Wako, Saitama 351-0198, Japan

[4]Department of Physics, McCullough Building, Stanford University, Stanford, California 94305-4045, USA

[#] These authors contribute equally to this work

* dingzhang@mail.tsinghua.edu.cn

* yongxu@mail.tsinghua.edu.cn

* qkxue@mail.tsinghua.edu.cn

## Abstract

**A single atomic slice of α-tin—stanene—has been predicted to host quantum spin Hall effect at room temperature, offering an ideal platform to study low-dimensional and topological physics. While recent research has intensively focused on monolayer stanene, the quantum size effect in few-layer stanene could profoundly change material properties, but remains unexplored. By exploring the layer degree of freedom, we unexpectedly discover superconductivity in few-layer stanene down to a bilayer grown on PbTe, while bulk α-tin is not superconductive. Through substrate engineering, we further realize a transition from a single-band to a two-band superconductor with a doubling of the transition temperature. *In-situ* angle resolved photoemission spectroscopy (ARPES) together with first-principles calculations elucidate the corresponding band structure. Interestingly, the theory also indicates the existence of a topologically nontrivial band. Our experimental findings open up novel strategies for constructing two-dimensional topological superconductors.**

Confining superconductivity to a two-dimensional (2D) plane engenders a variety of quantum phenomena [1,2]. Of late, the realization of highly crystalline and atomically thin superconductors has triggered a flurry of discoveries including the Griffiths singularity behavior [3], a quantum metallic phase [4,5], as well as an extremely large critical magnetic field in the plane [6,7,8]. One strategy of achieving 2D superconductors is to epitaxially grow superconductive single elements, such as Pb, In and Ga, for just one or two atomic layers [3,9,10]. Among the single elements, tin (Sn) is the very material in which the Meissner effect was first discovered [11] but realizing ultrathin Sn in the superconductive β-phase, known as white tin [12], remains challenging. The epitaxially grown Sn in the ultrathin limit tends to fall in the α-phase instead [13], whose bulk is semi-metallic and non-superconductive.

Recently, however, intensive research has been devoted to investigate the thinnest possible slice of α-tin (111)—a counterpart of graphene called stanene [14]. Stanene promises various exotic features such as highly efficient thermoelectrics [15], topological superconductivity [16], high-temperature quantum spin Hall [17] and quantum anomalous Hall effects [18]. Monolayer stanene that has been successfully fabricated by molecular beam epitaxy (MBE) on $Bi_2Te_3$(111) [19] and PbTe(111) [20] is the focus of current research. On the other hand, few-layer stanene is expected to show significant thickness-dependent properties due to the strong quantum confinement [21] but its exploration is still lacking.

In this Letter, by going from a monolayer to few-layer stanene, surprisingly, we discover superconductivity. We report the stable superconducting properties of uncapped few-layer stanene films on PbTe (111)/$Bi_2Te_3$ substrates. The superconducting transition temperature ($T_c$) can be effectively enhanced by varying the thickness of the PbTe buffer layer. Concomitantly with a doubling of $T_c$, we observe a single-band to two-band transition, which is further elucidated by photoemission spectroscopy and theoretical calculations. The calculated band structure further indicates the existence of inverted bands in our system. Our results therefore underscore the potential of an in-plane integration of two-dimensional topological insulator and superconductor—of the same material. The heterostructure, vertically consisting of superconducting few-layer stanene and topological insulator (TI) $Bi_2Te_3$, may also be of interest for inducing topological superconductivity via proximity effect [22].

Figure 1 **a** schematically illustrates the sandwich structure of our system with a trilayer Sn on top of PbTe/$Bi_2Te_3$/Si (111). The α-phase of Sn is confirmed by *in-situ* structural analysis (see Extended Data Figure 1). The dangling bonds on the top surface (Fig. 1 **a**) are presumably saturated, which is evidenced by our APRES data showing large band gaps at the K/K' points [19,20]. The saturation might be caused by hydrogen, a ubiquitous residue in the crystal growth environment [23], resulting in chemically stable samples. Figure 1 **b** shows that superconductivity emerges starting from a bilayer. By increasing the number of Sn layers ($N_{Sn}$), the transition temperature is consecutively promoted. In general, $T_c$ scales with $1/N_{Sn}$ (Fig. 1 **d**), as has been seen previously in other ultrathin films [1, 2]. We confirm the Meissner effect by a two-coil mutual inductance technique in Extended Data Figure 2. Extended Data Figure 3 further reveals the 2D nature of such a superconductor, evidenced by anisotropic critical magnetic fields and the Berezinskii-Kosterlitz-Thouless transition. As shown in Fig. 1 **c**, superconductivity also depends keenly on the thickness of the PbTe layer ($N_{PbTe}$). It emerges at $N_{PbTe}$=6 and $T_c$ further doubles when $N_{PbTe}$ exceeds 8. We speculate that this evolution stems from the change in density of states as well as the release of strain from the lattice mismatch (see Methods) [20]. A thicker PbTe might host more surface vacancies due to the lowered formation energy [24, 25], thus providing more electron doping into Sn as we will reveal by ARPES later. Notably, the superconductivity in these uncapped samples barely changes after exposing to air, as exemplified by the data taken in the second cool-down after more than two weeks of storage (Fig. 1 **c**). Extended Data Figure 2 further documents the superconductivity after one year of storage. In contrast, previous *ex-situ* transport studies on ultrathin Pb, In, and Ga films all rely on capping with an additional layer of Au or Ag [1-3]. We also show traces from two pairs of samples with the same nominal thicknesses, attesting to precise growth control. The transition temperature $T_c$ as a function of $N_{PbTe}$ is given in Fig. 1 **e**. The shaded regions represent two regimes corresponding to samples with $T_c$~0.5 K and those with $T_c$~1.2 K.

Transport properties of trilayer stanene in regime-I and II are distinctly different. For 3-Sn/10-PbTe in regime-II, its critical current with increasing temperature displays two steps with a kink at about 0.5 K (Fig. 2 **a**)—a characteristic feature of two-band superconductivity [26]. Such two-band nature is further confirmed by the temperature dependence of the upper critical field [27, 28]. The 3-Sn/10-PbTe sample displays a concave function of $\mu_0 H_c(T)$ (Fig. 2 **b**), which can be fitted by a formula designated to the two-band situation [27] (see Methods). In contrast to the behaviors of samples in regime-II, we observe no deviation from a single band superconductor for samples in regime-I to the lowest attainable temperature. Furthermore, they show different activated behaviors in the presence of a magnetic field. For 3-Sn/8-PbTe, fittings to the activated region extrapolate to a fixed point: $1/T_c$. In contrast, 3-Sn/10-PbTe displays a continuous shift of the crossing between the adjacent extrapolated lines (dashed in Fig. 2 **d**). The distinction is better captured by the extracted activation energy $U_H$ and the intercept of the fitting ln$R_0(\mu_0 H)$. In regime-I, $U_H$ scales linearly with $\ln(\mu_0 H)$ which can be described by the collective creeping of vortices [4] and the slope yields a London penetration depth $\Lambda$ of 700 nm (see Methods). The Ginzburg-Landau parameter $\kappa = \Lambda/\xi$ is therefore about 23, which is much larger than $1/\sqrt{2}$, as expected for a type II superconductor. In regime-II, one obtains instead a convex dependence of both $U_H$ on

$\ln(\mu_0 H)$ (Fig. 2 **e**) and $\ln R_0$ on $U_H/k_B T_c$ (Fig. 2 **f**). This nonlinearity may stem from field dependent superconducting parameters of $d_{sc}$ and $\Lambda$ for multiband superconductors [29, 30].

The transition from a single-band to two-band system is corroborated by APRES. Figure 3 **a** displays the data of a trilayer stanene with increasing $N_{PbTe}$. Two valence bands can be identified: a parabolic band with its highest intensity (dark color) below the Fermi level ($E_F$) and a linearly-dispersed band (white) with its two arms crossing $E_F$. The position of the Fermi level is distinctly different from that of bulk α-Sn [31, 32]. Superconductivity in few-layer stanene here may therefore stem from the enhanced density of states. By increasing $N_{PbTe}$, the two valence bands sink down, evidenced by the decrease in energy (Fig. 3 **b**) of the parabolic band as well as the shrinking Fermi momentum for the linear band (Fig. 3 **c**). They indicate an increase of electron transfer from PbTe. Concomitantly, a third band becomes discernible at the Fermi level. We focus on the region just below the Fermi level in the momentum range of [-0.2, 0.2] Å$^{-1}$. The overall shape evolves from a rounded pyramid for 3-Sn/6-PbTe to an hour-glass structure for $N_{PbTe} \geq 10$. Such an evolution is in direct contrast to the monotonic behavior of the residual photoelectron intensities in the gapped region, as seen in SrTiO$_3$ due to correlation effects [33]. We therefore attribute the hour-glass feature to the emergence of an electron pocket around the Γ. In the case of 3-Sn/6-PbTe, this electron pocket may just touch the Fermi level, providing negligible contribution in transport. With further doping, the central electron pocket gets significantly enlarged while the outer linear band shrinks. The trilayer stanene on PbTe with $N_{PbTe} \geq 10$ therefore behaves as a two-band superconductor. At higher doping, the enhanced interband scattering may suppress superconductivity [34], thus explaining the drop of $T_c$ for 3-Sn on 20-PbTe (Fig. 1 **c** and **e**). In addition, we estimate electron-phonon coupling constant to be 0.5±0.2 for the hole band [35], which agrees with our transport result by fitting the upper critical field data (Fig. 2 **b**). In comparison, for the bulk β-Sn $\lambda \sim 0.7$ [12].

We also performed first-principles calculation for a trilayer-stanene grown on PbTe (see Methods). The calculated band structure, displayed in Fig. 4, looks somewhat complicated, since orbitals of stanene and PbTe hybridize strongly with each other, showing significant Rashba splittings. Nevertheless, there exists two series of valence bands mainly contributed by stanene located about 0-0.6 eV below the valence band maximum (VBM). Importantly, the top valence bands are "M" shaped, which could introduce an "electron pocket" centered at Γ if placing $E_F$ slightly below the VBM. These features echo with the ARPES data. Furthermore, orbital analysis shows that Sn-*s* (Sn-*p*) orbitals have significant contribution to the lowest conduction (highest valence) band, except at Γ where an *s-p* band inversion happens. This band inversion results in a topologically non-trivial phase [17, 21, 36]. The trilayer stanene grown on PbTe is therefore a 2D TI with $Z_2=1$ in theory (see Extended Data Figure 5). We note that a band inversion may be induced in $Pb_{1-x}Sn_x Te$ alloy by increasing Sn, which is accompanied by reopening of the bulk band gap [25, 37, 38]. Experimentally, we observed no bulk band gap closing in PbTe with the low temperature deposition of Sn [20], ruling out possible topological transition in the PbTe substrate.

The delicate dependence of $T_c$ on $N_{Sn}$ can be employed for an in-plane integration of topological insulator and superconductor in the same material with tunable properties. Another direction for future endeavor is to investigate the proximity effect in the vertical direction. Our sandwich structure allows atomically sharp interfaces between a superconductor, a tunable barrier and a topological insulator—$Bi_2Te_3$. The Fermi momentum of few-layer stanene is comparable to that of $Bi_2Te_3$. Furthermore, the superconducting thickness we estimated can be larger than the total thickness of Sn and PbTe layers (Extended Data Figure 3 **d**) such that Cooper pairs may travel into $Bi_2Te_3$. In addition, stanene is robust against air exposure and can protect the more sensitive $Bi_2Te_3$. In general, the observation of superconductivity in few-layer stanene enriches the material pool for constructing topological devices.

**Data availability**

The data that support the findings of this study are available from the authors upon reasonable request.

## Acknowledgements

We thank Hong Yao, Canli Song for useful discussions. This work is financially supported by the Ministry of Science and Technology of China (2017YFA0304600, 2017YFA0302902) and the National Natural Science Foundation of China (grant No. 11604176). Y.X. acknowledges support from Tsinghua University Initiative Scientific Research Program and the National Thousand-Young-Talents Program. S.-C. Z. is supported by the U.S. Department of Energy, Office of Basic Energy Sciences, Division of Materials Sciences and Engineering under Contract No. DE-AC02-76SF00515.

## Author contributions

M. L. and Y. Z. contributed equally to this work. D. Z., K. H. and Q.-K. X. conceived the project. Y. Z. grew the samples and carried out ARPES measurements with the assistance of Y. G.. M. L. and D. Z. carried out the transport measurements with the assistance of K. Z.. M. L., D. Z., H. L., X. H., and Y.-Y. W. carried out the two-coil mutual inductance measurements.  Z. G. and Y. X. performed first-principles calculations.  D. Z. and Y. X. analyzed the data and wrote the paper with the input from K. H., X.-C. M., S.-C. Z. and Q.-K. X.. All authors discussed the results and commented on the manuscript.

## Competing financial interests

The authors declare no competing financial interests.

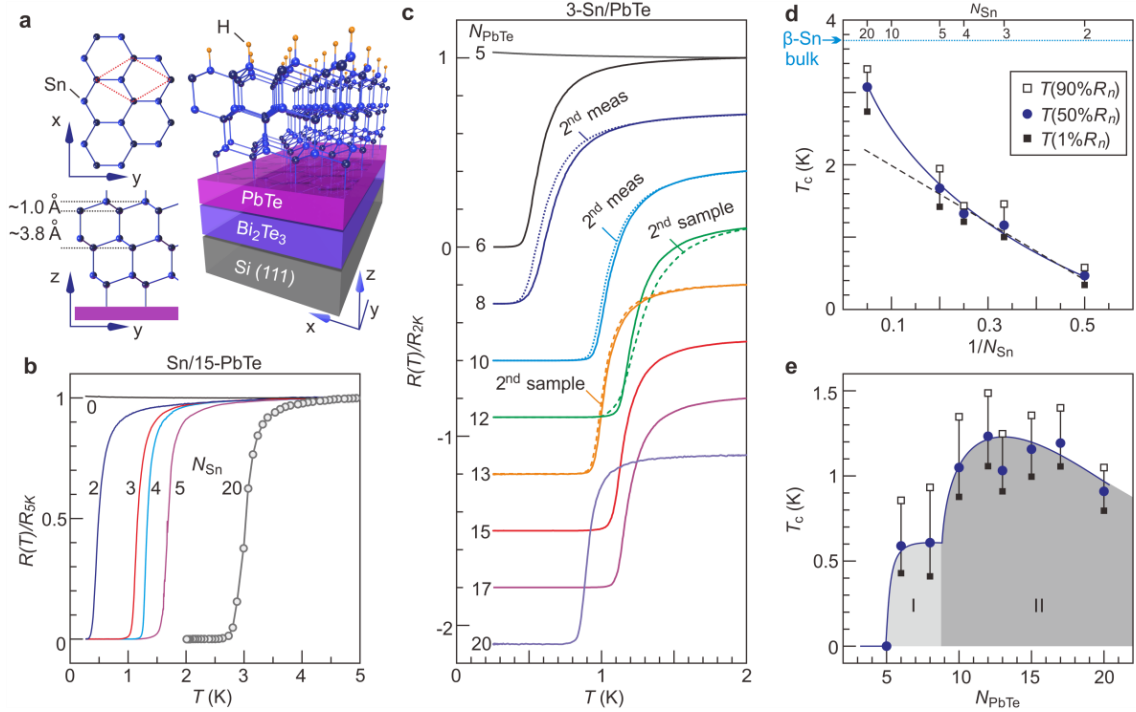

**Figure 1 Superconductive properties of few-layer stanene. a,** illustration of stanene lattice and the Sn/PbTe/Bi$_2$Te$_3$ sandwich structure. Upper left panel shows a top view of only one layer of stanene. Distances marked are from first-principle calculations. **b,** Normalized resistance of stanene with increasing number of layers grown on substrates consisting of 15-PbTe/5-Bi$_2$Te$_3$/Si(111). **c,** Normalized resistance of trilayer stanene (3-Sn) grown on different thicknesses of PbTe substrates. Numbers in the panel marks the number of PbTe layers. Dotted curves represent the data from the second measurement after 15-20 days of storage in a glovebox. Dashed curves are from a second sample grown in the same nominal thicknesses. Except for the top two, other curves are equally offset for clarity. **d e,** Critical temperature ($T_c$) as a function of the number of stanene layers (**d**) or the PbTe layers (**e**). The three data points of $T_c$ in a row represent the temperatures where the resistance drops to 1%, 50% and 90% of the normal resistance ($R_n$), respectively. Dashed straight line and solid curves in **d** and **e** are guide for the eye. For $N_{Sn}=20$, the superconducting transition temperature is approaching that of bulk β-Sn: 3.7K [12].

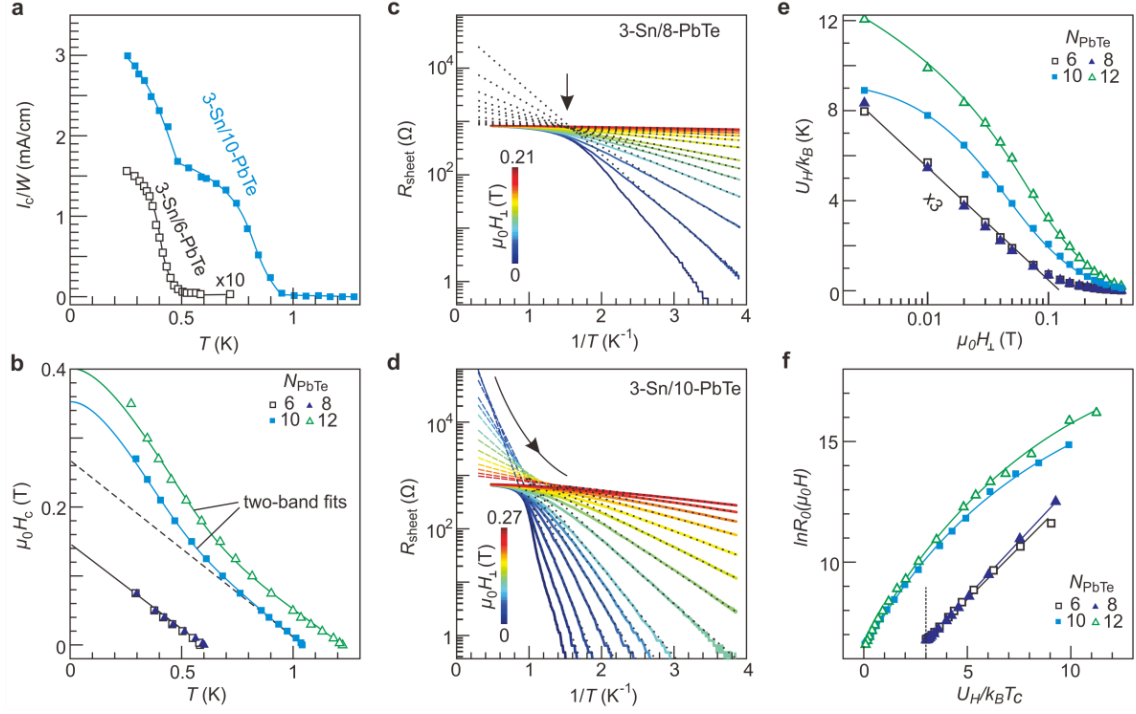

**Figure 2 Single band to two band transition of a trilayer stanene. a,** Critical current normalized by the width of sample. Curves are guide to the eye. **b,** Upper critical field as a function of temperature. From the two-band fitting, we obtain the ratio between the diffusivities of the two bands: $D_2/D_1 \sim 0.3$. The fitting also yields the electron-phonon coupling constants for the two respective bands and the interband one: $\lambda_{11} = 0.28$, $\lambda_{22} = 0.26$, $\lambda_{12} = 0.013$ (see Methods). **c d,** Arrhenius plots of the sheet resistance at different magnetic fields for two 3-Sn samples on different numbers of the PbTe layer. Dotted lines are linear fits to the data in the low temperature regime, reflecting the thermal activation behavior. **e,** Activation energy $U_H/k_B$ as a function of the perpendicular magnetic field. **f,** Extracted intercepts from fitting the activated transport (dotted lines in panel **c** and **d**). Data points for $N_{PbTe}=6$ and 8 are horizontally offset for clarity (dotted line marks the zero point).

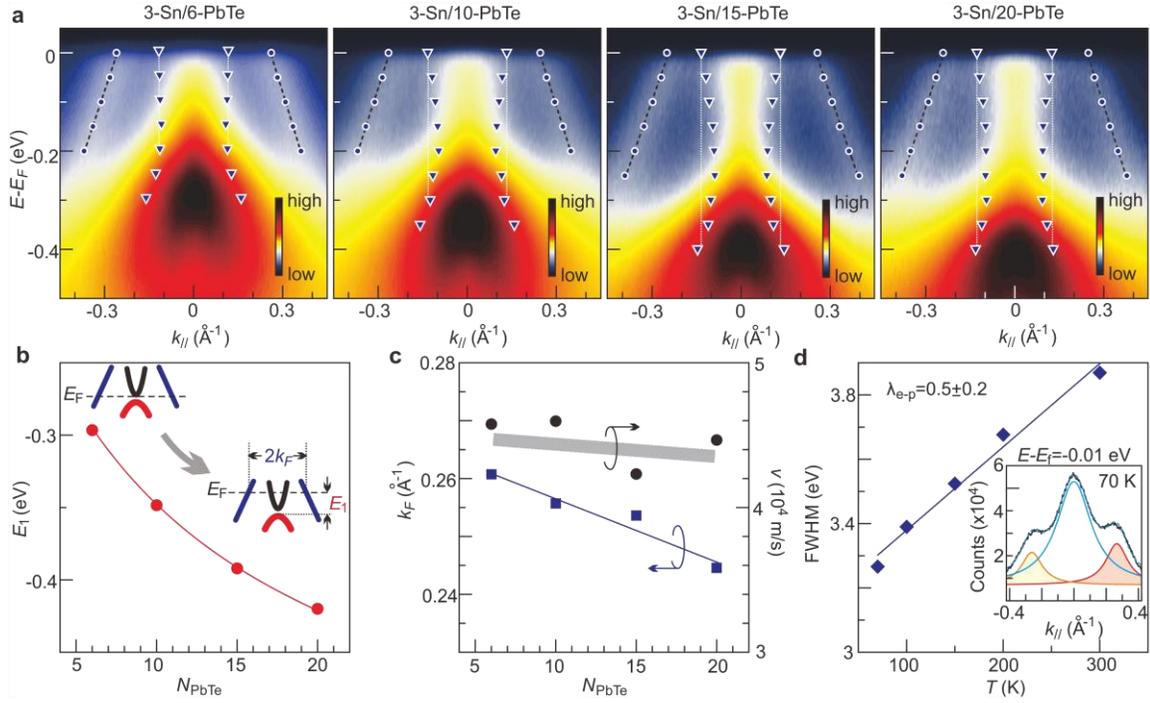

**Figure 3 ARPES studies of a trilayer stanene. a,** Band structure around the Γ point for 3-Sn grown on increasing number of PbTe layers. Circles mark the linear dispersion of the hole bands with dashed lines as linear fits. Triangles demarcate the width of the central peak around $k_{//}=0$ at different energies. Dotted vertical lines are guide to the eye. **b,** Downward shift of the energy band. Inset explains the definitions of $E_1$ and $k_F$. **c,** Fermi momentum ($k_F$) and the velocity of the linearly dispersed hole band as a function of $N_{PbTe}$. **d,** Energy width of the hole band as a function of temperature for a 3-Sn/10-PbTe sample. The electron-phonon coupling constant is estimated from a linear fitting (solid line) to the data points [35]. The estimated electron phonon coupling constant is 0.5±0.2 with the uncertainty obtained by taking into account both the Lorentzian and the linear fittings involved (see Extended Data Figure 4). Inset illustrates the Lorentzian fitting to the momentum distribution curve (MDC) at $E-E_F=-0.01$ eV, 70 K. The shaded peaks reflect the hole bands. The energy width is calculated from the product of the momentum width of the shaded peaks (Δk) and the slope of the band (dE/dk).

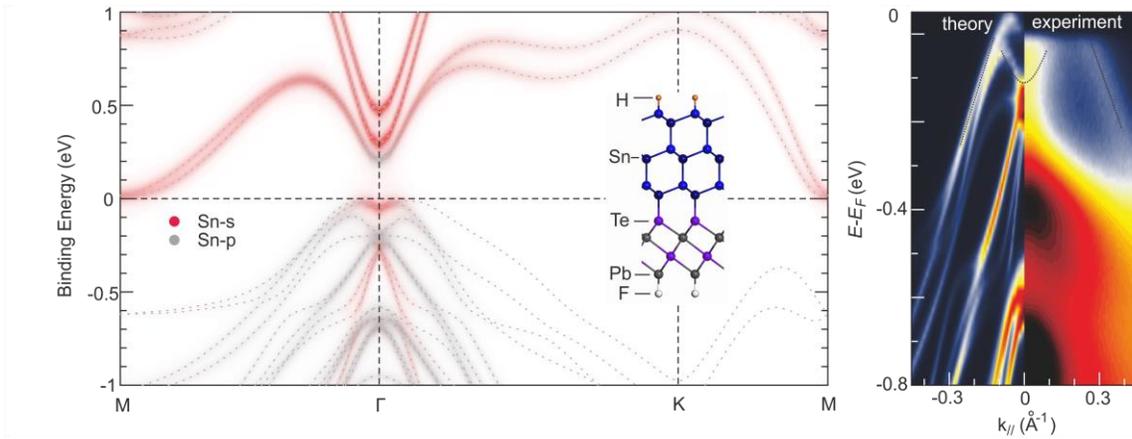

**Figure 4 Calculated band structure of a trilayer stanene on PbTe.** Inset illustrates the atomic model considered in the first-principles calculation. A hydrogenated trilayer stanene is placed on top of the PbTe substrate, which is simulated by a slab of two PbTe layers saturated by fluorine on the bottom. Red (gray) color highlights the contributions from Sn-*s* (Sn-*p*) orbital, obtained by projecting the Bloch wavefunction onto the corresponding orbitals. It indicates an *s-p* band inversion at the $\Gamma$ point. The right panel compares the calculated bands with the ARPES data from 3-Sn/15-PbTe.

**Methods**

**Growth:** We use molecular beam epitaxy to grow our heterostructures (Omicron, base pressure $1 \times 10^{-10}$ mbar). To ensure lattice matching, five quintuple layers of $Bi_2Te_3$ was first grown on top of Si(111) substrates. This was followed by the layer-by-layer growth of PbTe. Finally, we deposit Sn at a substrate temperature of around 120 K. The sample is then annealed at temperatures up to 400 K to improve the film quality. The crystalline quality is monitored by *in-situ* reflective high energy electron diffraction (RHEED) and scanning tunneling microscopy (STM) (see Extended Data Figure 6). A layer-by-layer growth is maintained from a monolayer up to the quintuple layer. Above five layers, the growth tend to form islands. The lattice constant of stanene expands as the number of PbTe layer increases, as revealed by RHEED [20].

**Transport:** Samples grown on intrinsic Si(111) substrate were employed for low temperature transport measurements in a closed cycle system (Oxford Instruments TelatronPT) equipped with a He-3 insert (base temperature=0.25 K). The temperature sensor was placed right below the sample stage and positioned in an orientation with minimal magnetoresistances. Freshly cut indium cubes were cold pressed onto the sample as contacts. Standard lock-in techniques were employed to determine the sample resistance in a four-terminal configuration with a typical excitation current of 100 nA at 13 Hz.

To fit upper critical field as a function of temperature in the two-band regime, we employ the formula [27]:

$$\ln\frac{T}{T_c} = -\frac{\left[U\left(\frac{eD_1\mu_0 H_{c2}}{hT}\right) + U\left(\frac{eD_2\mu_0 H_{c2}}{hT}\right) + \frac{\sqrt{(\lambda_{11}-\lambda_{22})^2 + 4\lambda_{12}^2}}{\lambda_{11}\lambda_{22}-\lambda_{12}^2}\right]}{2} + \left[\frac{\left(U\left(\frac{eD_1\mu_0 H_{c2}}{hT}\right) - U\left(\frac{eD_2\mu_0 H_{c2}}{hT}\right) - \frac{\lambda_{11}-\lambda_{22}}{\lambda_{11}\lambda_{22}-\lambda_{12}^2}\right)^2}{4} + \frac{\lambda_{12}^2}{(\lambda_{11}\lambda_{22}-\lambda_{12}^2)^2}\right]^{\frac{1}{2}},$$

where $U(x) = \psi\left(\frac{1}{2} + x\right) - \psi\left(\frac{1}{2}\right)$ with $\psi$ the digamma function. $D_1$ and $D_2$ reflect the diffusivities of the two bands. $\lambda_{11}$, $\lambda_{22}$, and $\lambda_{12}$ are intraband and interband electron-phonon coupling constants, respectively. We fit the data of 3-Sn/10-PbTe with $D_1$, $D_2$, $\lambda_{11}$, $\lambda_{22}$, and $\lambda_{12}$ as fitting parameters. We then use the extracted values of $\lambda_{11}$, $\lambda_{22}$, and $\lambda_{12}$ and fit the data of 3-Sn/12-PbTe with $D_1$ and $D_2$ as free parameters.

For the activated transport, we use $R_{sheet} = R_0(\mu_0 H) e^{-\frac{U_H}{T}}$ to fit the data. $U_H$ represents the activation energy. In regime-I, $R_0(\mu_0 H)$ scales as $R_0 e^{\frac{U_H}{T_c}}$, with $R_0$ being independent of $\mu_0 H$. Also, $U_H$ scales linearly with $\ln \mu_0 H$ such that: $-\frac{dU_H}{d \ln \mu_0 H} = \frac{\Phi_0^2 d}{256 \pi^3 \Lambda^2}$ [39]. Here $\Phi_0$ is the flux quantum and $\Lambda$ the London penetration depth normal to the superconducting film.

**ARPES:** Samples grown on highly doped Si(111) substrates were transferred to the analysis chamber without breaking the ultrahigh vacuum. ARPES with a photon energy of 21.22 eV (He-I light) were carried out with a Scienta R4000 spectrometer. For a quantitative analysis, we first extract the momentum $k$ at a series of binding energies by fitting the peaks in the corresponding momentum distribution curves (MDC) with Lorentzian functions. The obtained data points: $k(E)$ (white circles in Fig.3 **a**) are then linearly fitted (dashed lines) to extract $dE/dk$ as well as $k_F$.

**First-principles calculations:** Density functional theory calculations were performed by the Vienna *ab initio* simulation package, using the projector-augmented-wave potential, the Perdew-Burke-Ernzerhof exchange-correlation functional and the planewave basis with an energy cutoff of 400 eV. The periodic slab approach was employed to model stanene grown on PbTe, using a vacuum layer of 12 Angstrom and a 12×12×1 Monkhorst-Pack $k$ grid. A slab of two Pb-Te bilayers with a surface lattice constant of 4.568 Angstrom (based on the experimental value of bulk) was used to simulate the substrate, in which the bottom Pb-Te bilayer was fixed during relaxation and the bottom Pb atoms were saturated by fluorine for removing the dangling bonds on the bottom. The spin-orbit coupling was included in the self-consistent calculations of electronic structure.

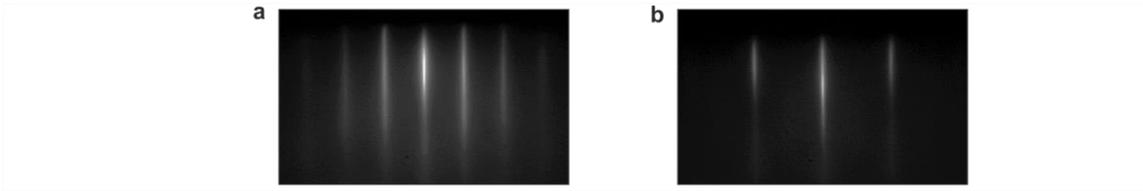

**Extended Data Figure 1 Reflection high-energy electron diffraction (RHEED) patterns demonstrating the triangular lattice of epitaxial α-Sn thin film.** The sample consists of 2-Sn (bilayer stanene) grown on 10-PbTe. **a** and **b** represent RHEED patterns from [1$\bar{1}$0] and [11$\bar{2}$] directions. Sharp and clear stripes demonstrate layer-by-layer growth of the material. The separations of the stripes in **a** and **b** give rise to a ratio of $1:\sqrt{3}$, suggesting the sixfold symmetry. The estimated lattice constant is a=4.52 Å, in agreement with the lattice constant of α-Sn [19].

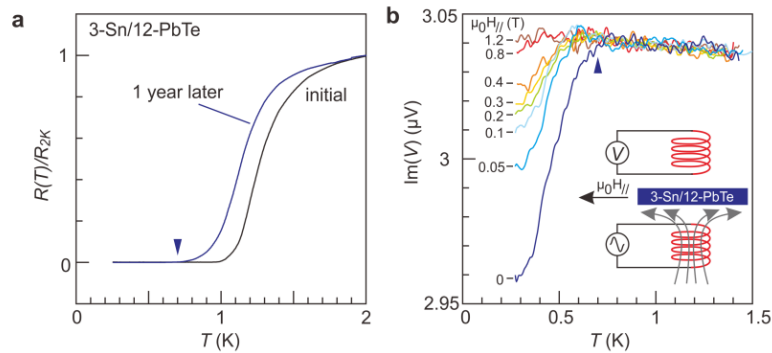

**Extended Data Figure 2 Meissner effect of a trilayer stanene. a,** Normalized resistance of a 3-Sn/12-PbTe sample measured just after the growth (initial) and after one year of storage in a glovebox. The sample remains superconductive after such an extended time of storage, demonstrating its robustness. **b,** Two-coil mutual inductance measurement on the 3-Sn/12-PbTe sample (after 1 year of storage). Inset sketches the measurement principle as well as the direction of the DC magnetic field. The superconducting sample screens the AC magnetic field generated by the drive coil (bottom). Consequently, the signal picked up by the second coil on the other side drops [40]. Blue arrows in **a** and **b** indicate the same $T_c$ of 0.7 K. As shown in panel **b**, the drop of the signal is completely suppressed at a strong in-plane DC magnetic field of about 1.2 T, confirming the large in-plane upper critical field for the 2D superconductor. This in-plane field experiment rules out the possibility of other spurious signals from the solder or indium residues, which have small upper critical fields. In our setup, each of the coils is made of 150 to 200 turns of thin insulated copper wires (30 μm in diameter with 25 μm of Cu and 5 μm of insulation layer) wound on a ceramic rod with a diameter of 1 mm. We applied 300 μA, 45 kHz to the drive coil (bottom in the inset) and connect the pick-up coil (top) to the lock-in system (Stanford Research SR-830).

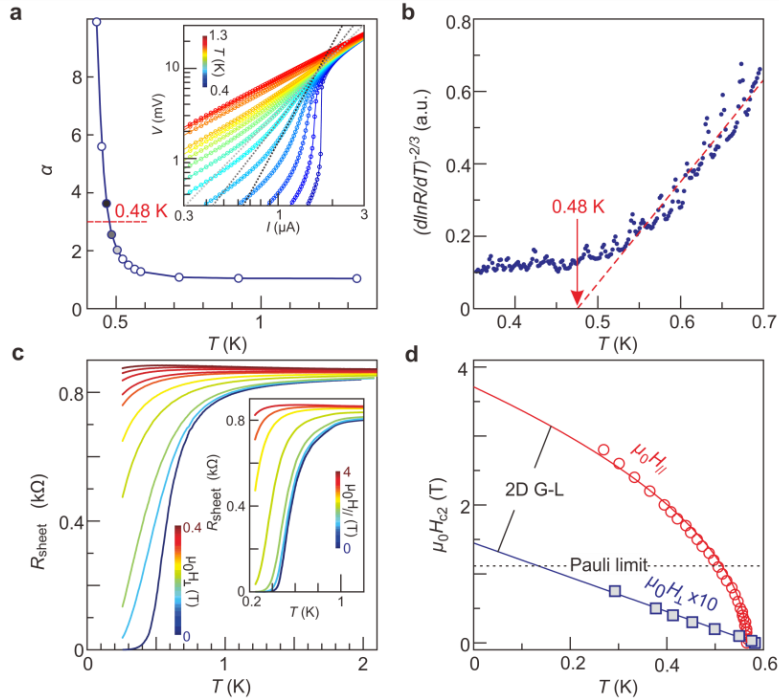

**Extended Data Figure 3 Evidence for a 2D superconducting transition.** The superconducting transition in a 2D superconductor is expected to be accompanied by an exponential suppression of unbound vortices and anti-vortices—a Berezinskii-Kosterlitz-Thouless (BKT) transition. Evidence for such a phase transition in our few-layer stanene is collected from (**a**) the dc *I-V* measurement and (**b**) the resistance curves of a 3-Sn on the 6-PbTe/5-Bi$_2$Te$_3$ substrate. With decreasing temperatures, a nonlinear behavior starts to replace the Ohmic response of the *I-V* dependence [41]. In the BKT theory, such a nonlinearity can be described as $V=I^\alpha$. The exponent α crosses 3 at about 0.48 K for 3-Sn/6-PbTe. **b**, $[d\ln(R)/dT]^{-2/3}$ as a function of temperature. Extrapolating the linear part to zero yields the BKT transition temperature of 0.48 K, consistent with the *I-V* measurement. **c,** Sheet resistance of a 3-Sn on the 6-PbTe/5-Bi$_2$Te$_3$ substrate at selected sets of magnetic fields applied either perpendicular or parallel to the film ($\mu_0 H_\perp$ =0, 0.02, 0.04, 0.1, 0.15, 0.21, 0.24, 0.3, 0.35, and 0.4 T, $\mu_0 H_\parallel$ =0, 0.5, 1, 2, 3, 3.5, and 4 T). **d,** Upper critical fields for 3-Sn/6-PbTe. By fitting to the 2D Ginzburg-Landau (GL) formula [4], we obtain an in-plane coherence length of $\xi_{GL}$ =30 nm and a superconducting

thickness of $d_{sc}$=6 nm. Dotted line indicates that the in-plane critical field well exceeds the Pauli limit.

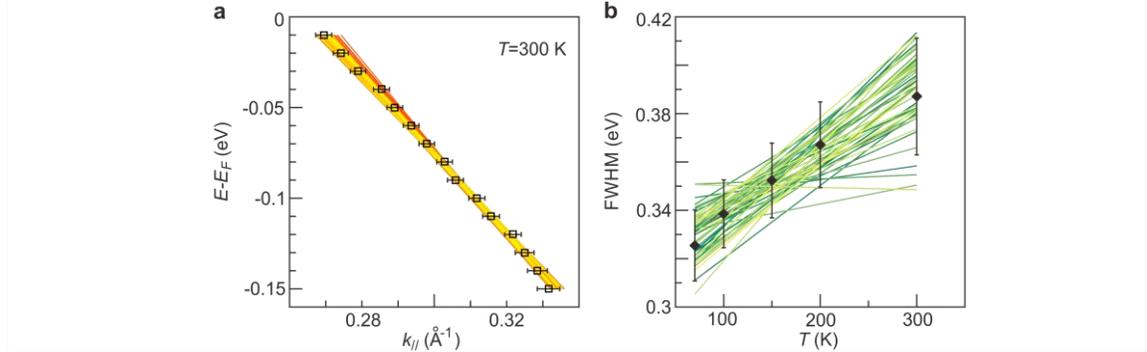

**Extended Data Figure 4 Estimation of error in the evaluated electron-phonon coupling constant. a,** Estimation of the error in $\frac{dE}{dk}$. Black squares indicate the relation between energy and momentum for the linearly dispersed hole band at 300 K. Here the error bars are from the Lorentzian fitting (inset of Fig. 3 **d**). Straight lines are 50 linear fits to the data points randomly distributed around the fitted momenta with a standard deviation marked by the error bar. These random numbers follow a Gaussian distribution. We carried out 10000 such linear fittings and estimated the standard deviation of the fitted slopes, which is taken as the error for the slope: $\text{err}\left(\frac{dE}{dk}\right)$. **b,** Estimation of the error in $\frac{dFWHM}{dT}$. Black diamonds are the estimated $FWHM$ at each temperature. The error bar at each temperature is calculated from: $\text{err}(FWHM) = \sqrt{(\Delta k)^2 \times \left(\text{err}\left(\frac{dE}{dk}\right)\right)^2 + \left(\frac{dE}{dk}\right)^2 \times (\text{err}(\Delta k))^2}$, where $\Delta k$ is the width of the peak in the MDC at -0.01 eV and $\text{err}(\Delta k)$ is the error from the Lorentzian fitting (inset of Fig. 3 **d**). Straight lines are 50 linear fits to the data points randomly distributed with a standard deviation of $\text{err}(FWHM)$ at each $T$. These random numbers follow a Gaussian distribution around $FWHM$. We again carried out 10000 such linear fittings and estimated the standard deviation of the fitted slopes, which yields: $\text{err}\left(\frac{dFWHM}{dT}\right)$. We then used this error to calculate the error of the electron-phonon coupling constant.

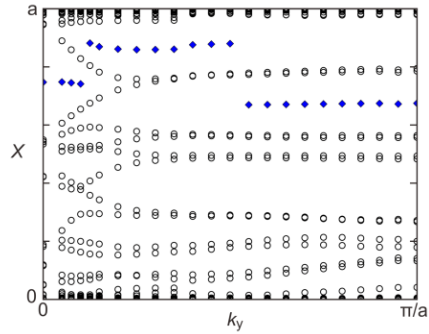

**Extended Data Figure 5 Evolution of the Wannier charge centers (circles) and their largest gap function (blue diamonds) for a trilayer stanene on PbTe.** The calculation, performed by the *Z2pack* code [42], gives $Z_2 = 1$, showing that the system is a 2D topological insulator.

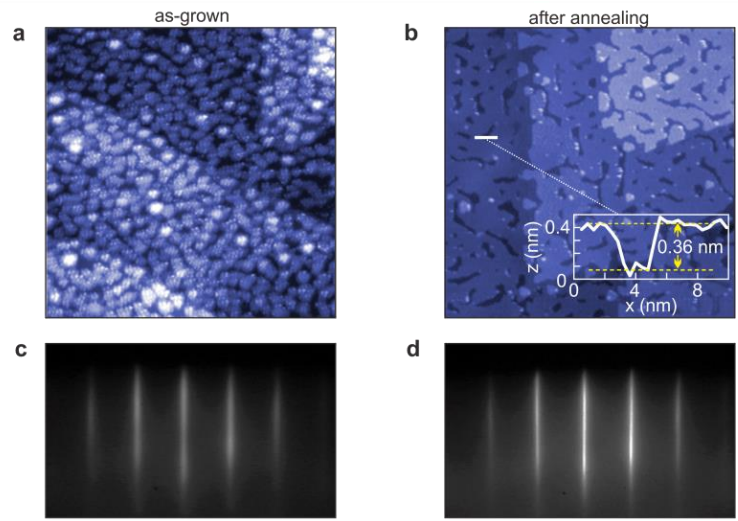

**Extended Data Figure 6 Comparison of the sample quality before and after annealing.** **a** and **b** (**c** and **d**) are STM (RHEED) images taken for the as-grown stanene (1-Sn/10-PbTe) and the sample after 2 hours of annealing at 373 K, respectively. The STM images are 100×100 nm². Inset in **b** provides a z-direction profile along the indicated white line. Here the bottom of the groove is the PbTe substrate. The step height (spacing between the dashed lines) is 0.36 nm. This value corresponds to the vertical distance from the top of the monolayer stanene to the PbTe substrate, consistent with the previous report [19].